\documentclass[epj-ND]{svjour}
\usepackage[tbtags]{amsmath}
\usepackage{amssymb}
\usepackage{graphicx}

\usepackage{txfonts}

\begin{document}

\sloppy

\title{A fully microscopical simulation of nuclear collisions
by a new QMD model}

\author{Maria Vittoria Garzelli \inst{} \thanks{Talk at ND2007, 
International Conference on
Nuclear Data for Science and Technology, April 22 - 27 2007, Nice, France.
Presenting author
\email{maria.garzelli@mi.infn.it}.  
}}

\institute{
University of Milano, Department of Physics, and INFN, Sezione di Milano,
I-20133 Milano, Italy
}

\abstract{
Nucleon-ion and ion-ion collisions 
at non relativistic bombarding energies can
be described by means of Monte Carlo approaches, such as those based
on the Quantum Molecular Dynamics (QMD) model. We have developed a
QMD code, to simulate the fast stage of heavy-ion reactions, and
we have coupled it to the de-excitation module available in the FLUKA
Monte Carlo transport and interaction code.
The results presented in this work span
the projectile bombarding energy range within 200 - 600 MeV/A, allowing to
investigate the capabilities and limits of our non-relativistic QMD approach.}

\maketitle

\section{General framework and motivation of this work}
\begin{figure}[htb!]
\centering
\vspace*{-9mm}
\resizebox{0.88\columnwidth}{!}{%
   \includegraphics[bb=0 30 516 523, clip]{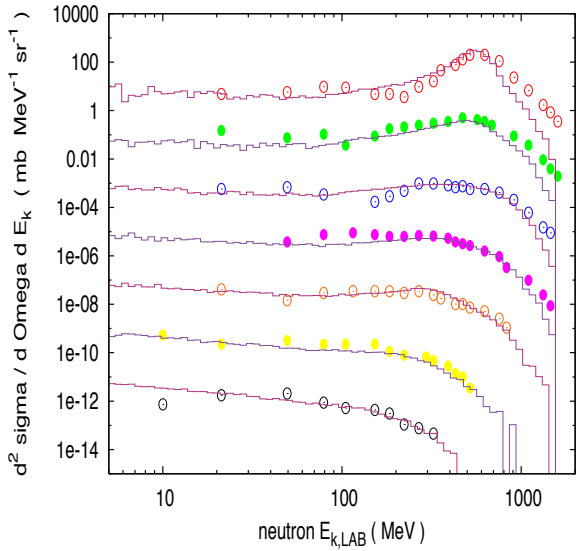}  
}
\caption{Double-differential neutron production cross-section for
Ar projectiles impinging on C at  560 MeV/A bombarding energy. 
The results of the theoretical simulations made by QMD + FLUKA 
de-excitation are shown
by solid histograms, while the experimental data taken from ref.~\cite{iwata}
are shown by circles. Distributions at
5$^{\mathrm{o}}$, 10$^{\mathrm{o}}$, 20$^{\mathrm{o}}$, 
30$^{\mathrm{o}}$, 40$^{\mathrm{o}}$, 60$^{\mathrm{o}}$ and 
80$^{\mathrm{o}}$ (lab) angles
are scaled by decreasing even powers of 10.}
\label{fig:1}
\end{figure}
\label{intro}
A fully microscopical simulation of nucleon-ion and ion-ion
collisions, at nucleon level, can be performed, among several
different approaches~\cite{bou}, 
by means of Quan\-tum Mo\-le\-cu\-lar Dynamics 
(QMD) models~\cite{aiche}. They are dynamical mo\-dels which allow to study
the phase-space evolution of the projectile-target
colliding systems, from their initial mutual trajectory influence 
and eventually their overlap, depending on the impact parameter, 
to the compression phase, accompanied by
a temperature and  density increase, up to the following expansion
stage, characterized by the formation of hot excited fragments (pre-fragments).
The whole phase occurs on a time scale within a few hundreds fm/c,
depending on the size of the colliding systems and the 
bombarding energy, and is called the ``fast'' stage of the reaction.
\begin{figure}[htb!]
\centering
\resizebox{0.95\columnwidth}{!}{%
   \includegraphics[bb=50 50 400 280, clip]{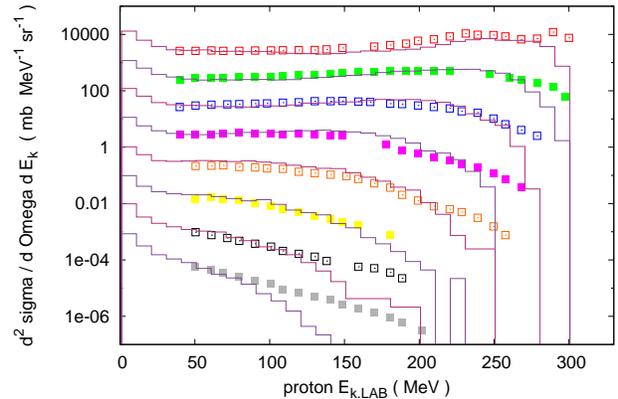}
}
\caption{Double-differential proton production cross-section for
p projectiles impinging on C at  300 MeV/A bombarding energy. 
The results of the theoretical simulations
are shown
by solid histograms, while the experimental data
from ref.~\protect\cite{proto}
are shown by squares. Distributions at
20$^{\mathrm{o}}$, 30$^{\mathrm{o}}$, 40$^{\mathrm{o}}$, 
50$^{\mathrm{o}}$, 60$^{\mathrm{o}}$, 75$^{\mathrm{o}}$, 90$^{\mathrm{o}}$  
and 105$^{\mathrm{o}}$ (lab) angles
are scaled by 
$10^{4}$, $10^{3}$, $10^{2}$, $10^{1}$, $10^{0}$, $10^{-1}$,
$10^{-2}$, $10^{-3}$, respectively.
}
\label{fig:2}
\end{figure}
   
Additionally, improved versions of QMD
models (e.g. the CoMD one, developed by Papa {\it et al.}~\cite{sici})
allow to compute the system evolution even for a longer time, up to
thousands fm/c, and thus have also been used to describe 
pre-fragment
de-excitation, at least in its initial stage. 
This has led to direct comparisons of the results of improved
QMD simulations to experimental data concerning fragment 
emission distributions.
On the other hand, 
pre-fragment
de-excitation can occur on
a time scale even larger (up to $\sim$ 10$^{-15}$ s). 
\begin{figure}[htb!]
\centering
\resizebox{0.95\columnwidth}{!}{%
   \includegraphics[bb=50 50 400 280, clip]{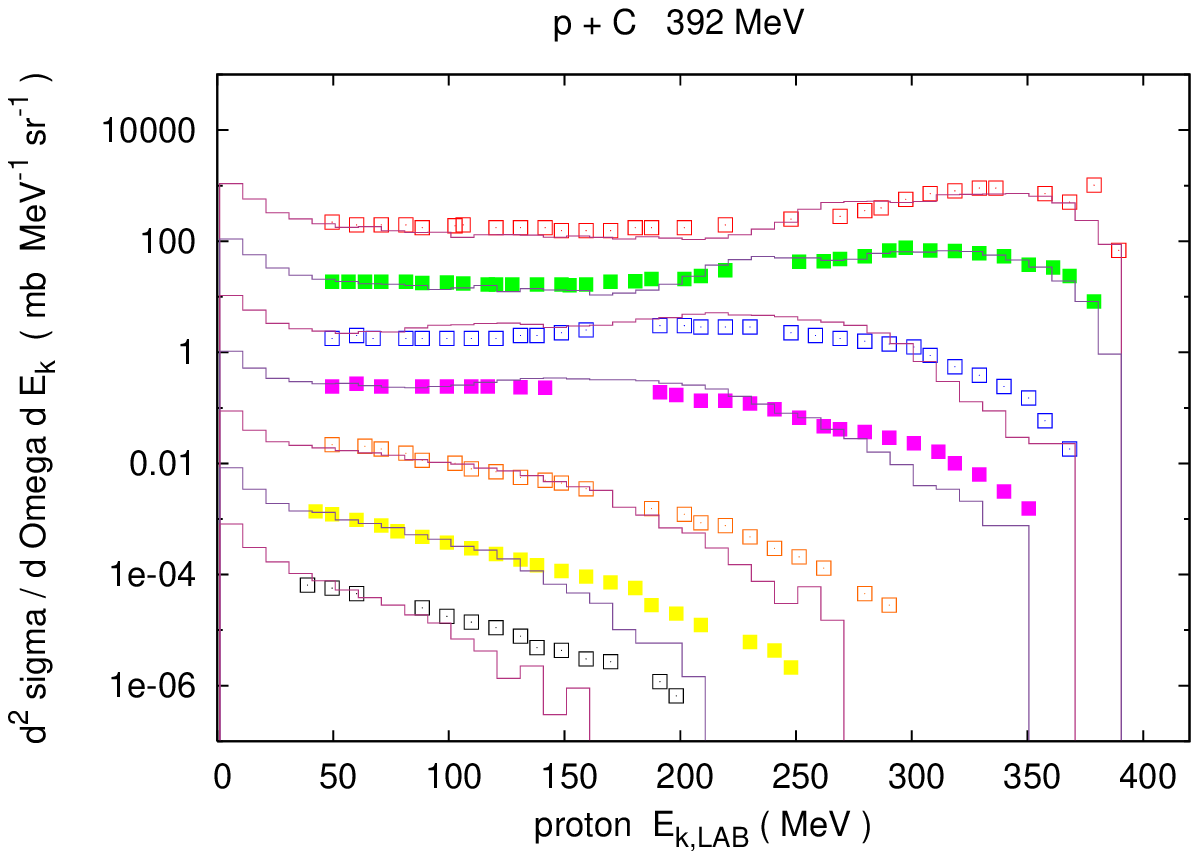}  
}
\caption{The same as fig.~\ref{fig:2} for p impinging on C at 392 MeV.
 Distributions at
20$^{\mathrm{o}}$, 25$^{\mathrm{o}}$, 40$^{\mathrm{o}}$, 
50$^{\mathrm{o}}$, 75$^{\mathrm{o}}$, 90$^{\mathrm{o}}$ and 
105$^{\mathrm{o}}$ (lab) angles
are scaled by
$10^{3}$, $10^{2}$, $10^{1}$, $10^{0}$, $10^{-1}$,
$10^{-2}$, $10^{-3}$, respectively.}
\label{fig:3}
\end{figure}
Thus, a complete treatment of this slow stage 
can be covered by different models,
generally based on statistical considerations.
The underlying assumption in applying 
one of these statistical models
to nuclear systems is that they are thermalized. 
While at the lowest energies the colliding ions stay close 
to each other for a time long enough for thermalization to occur,  
to define a temperature for the whole system 
at higher energies (several tens MeV/A)
can be very problematic, 
since the expansion phase can begin before a global thermalization process
is completed. Anyway, in the last case, 
at advanced time in the expansion stage 
pre-fragments are well separated and each of them is supposed
to be thermalized. Whereas theoretical models allow to compute 
an excitation energy for each pre-fragment,
from the experimental point of view
the problem of the determination of hot fragment
temperatures is still open.
On the other hand, planned applications (such as hadrontherapy
and space radioprotection) need models and tools to calculate
doses to human bodies and equipment, due to radiation exposure.
In particular, reliable calculations of physical doses require an accurate
description of nuclear interactions. Nuclear reaction models used for
predictions in these applications, 
should be capable of reproducing available experimental data 
concerning particle and fragment emission.
\begin{figure}[htb!]
\centering
\resizebox{0.815\columnwidth}{!}{%
   \includegraphics[bb=50 50 400 280, clip]{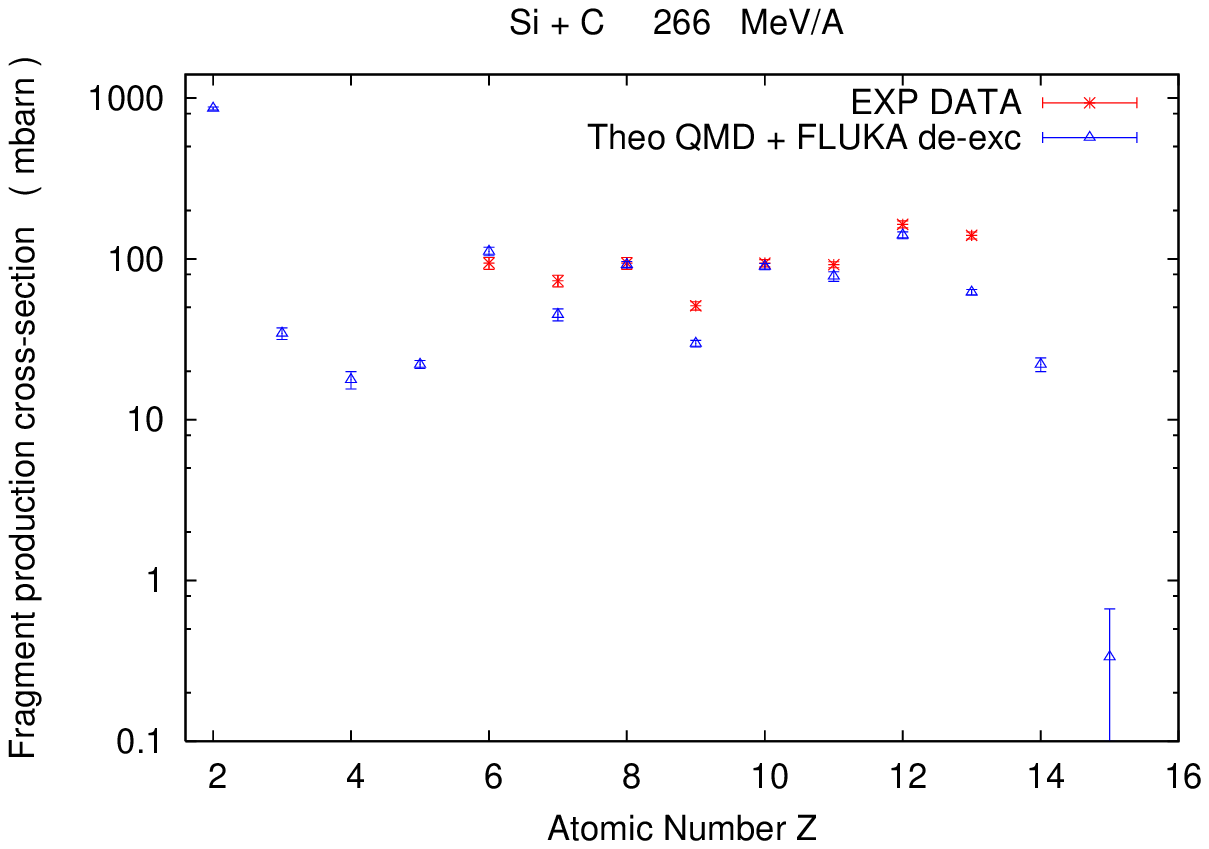}
}
\resizebox{0.815\columnwidth}{!}{%
   \includegraphics[bb=50 50 400 280, clip]{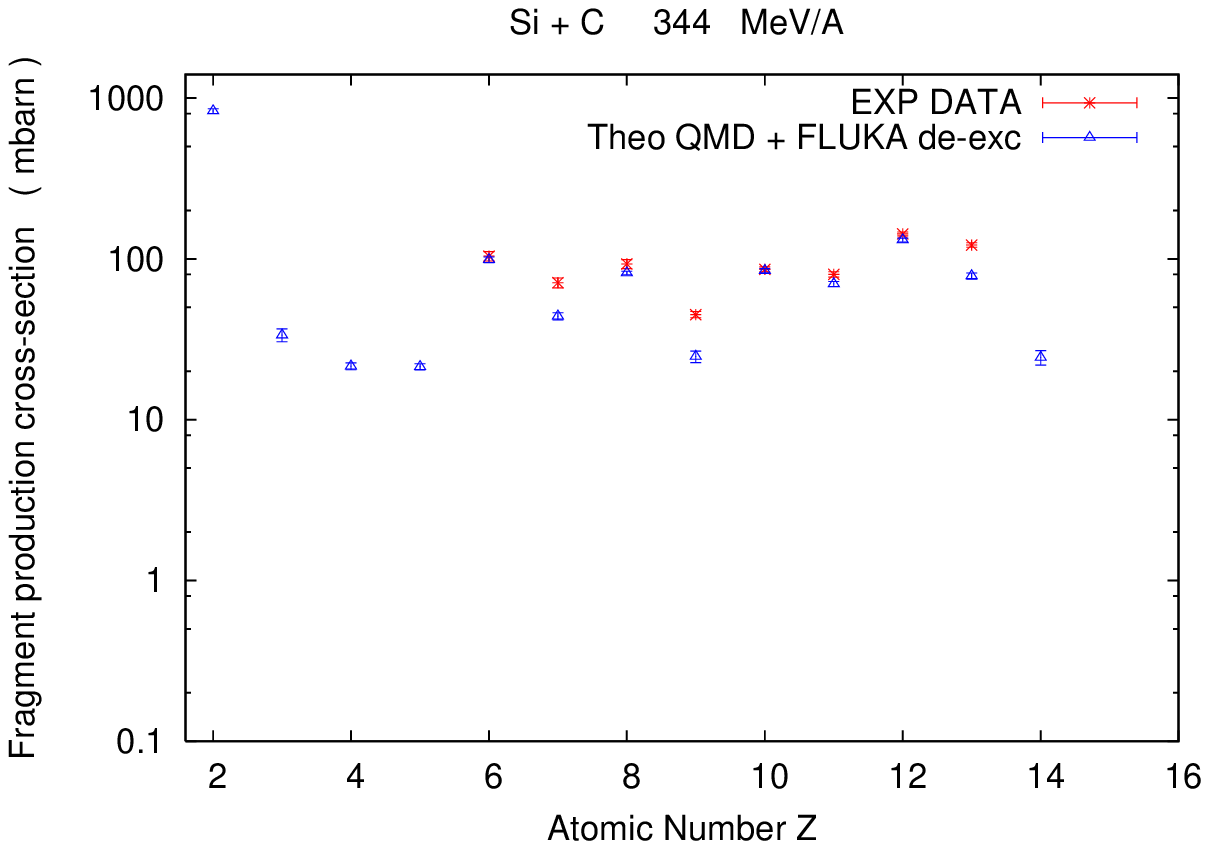}
}
\caption{Charged projectile-like fragment production cross-sections
for Si ions impinging on C at 266 and 344 MeV/A 
bombarding energies, respectively in the upper and in the lower panel: 
theoretical predictions 
are compared to experimental data from ref.~\cite{silicio}. 
Detector acceptance has been included
in the theoretical simulations.
\label{fig:4}}
\end{figure}

This paper presents the results of simulations performed with a
newly developed QMD code,
coupled to the de-excitation module available
in one of the most widely used Monte Carlo transport and interaction
codes, the FLUKA code~\cite{chep03,flukacern}. 
In particular, FLUKA includes algorithms allowing to compute evaporation, 
fission, fragmentation and Fermi break-up (for light nuclei)
followed by $\gamma$ emission, for whichever excited pre-fragment.
Those algorithms are part of the FLUKA general hadron-nucleus 
(and lepton-nucleus) interaction model called PEANUT (PreEquilibrium 
Approach to NUclear Thermalization)~\cite{trieste}. 
For further information, the interested reader can check also
refs.~\cite{nd2004,varennaalfredo,zfis1,zfis2} 
and the FLUKA website {\it http://www.fluka.org}.
The excited pre-fragments computed by the QMD model at the end of the
fast stage are then passed through these algorithms for the final
de-excitation and fragment generation.  
\section{Comparisons with experimental data}
\begin{figure}[htb!]
\centering
\resizebox{0.815\columnwidth}{!}{%
   \includegraphics[bb=50 50 400 280, clip]{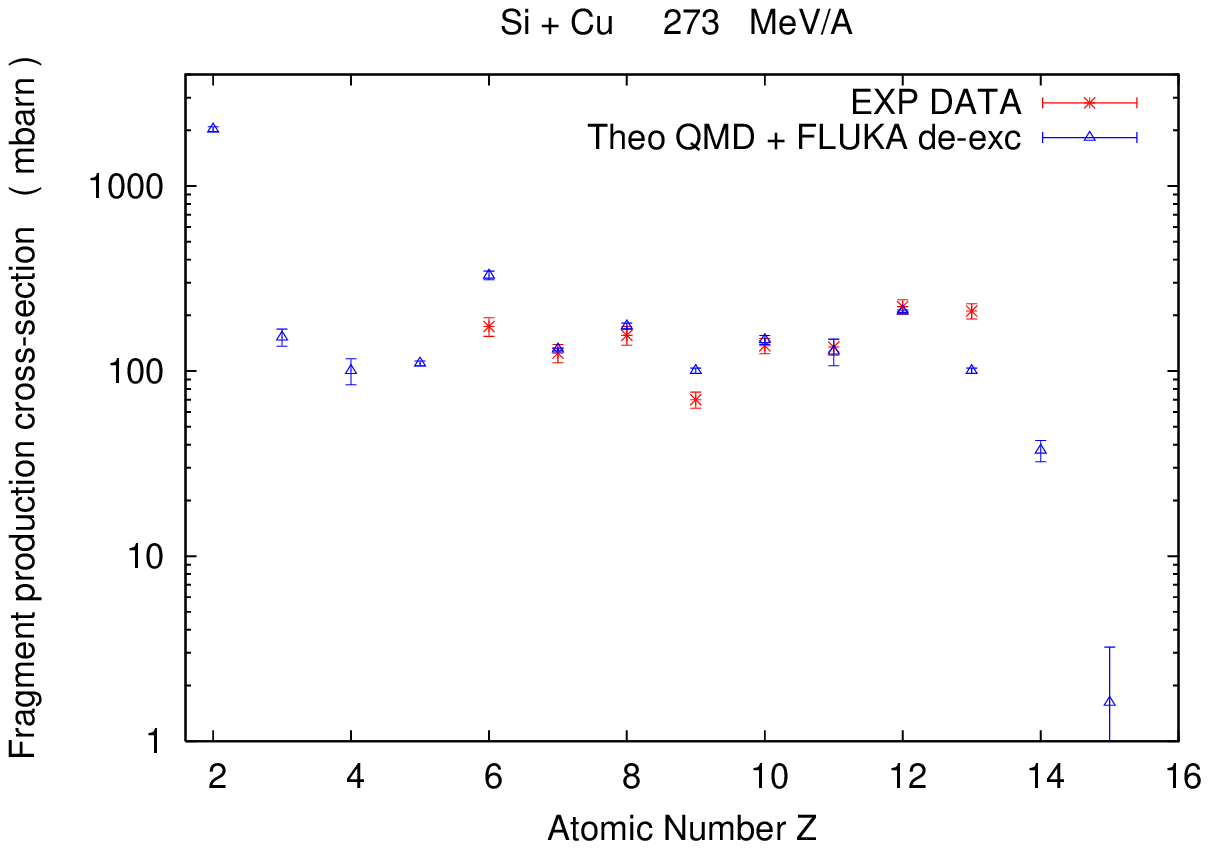}
}
\resizebox{0.815\columnwidth}{!}{%
   \includegraphics[bb=50 50 400 280, clip]{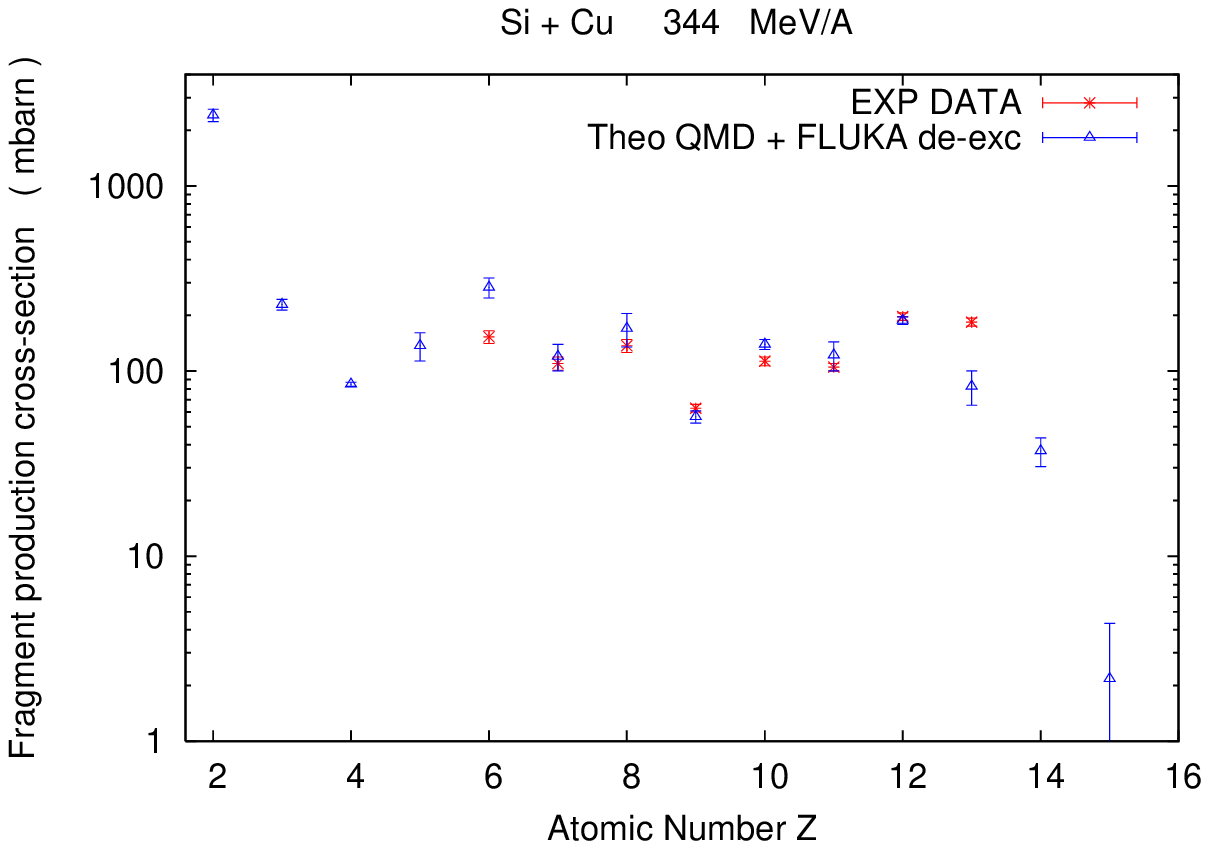}
}
\resizebox{0.815\columnwidth}{!}{%
   \includegraphics[bb=50 50 400 280, clip]{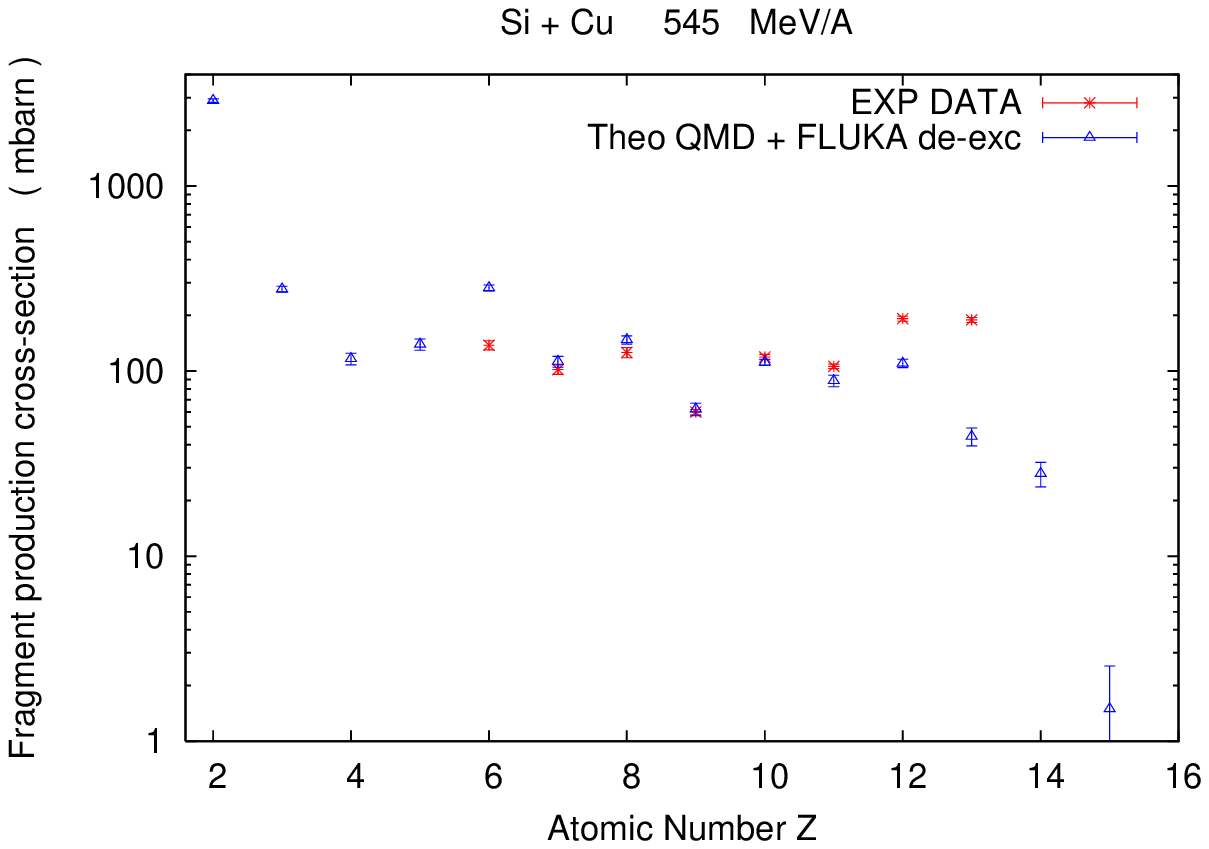}
}
\caption{The same as fig~\ref{fig:4}
for Si ions impinging on Cu at 274, 344 and 545 MeV/A 
bombarding energies, respectively in the upper, 
intermediate and lower panel.
\label{fig:5}}
\end{figure}
\begin{figure}[htb!]
\centering
\resizebox{0.815\columnwidth}{!}{%
   \includegraphics[bb=50 50 400 280, clip]{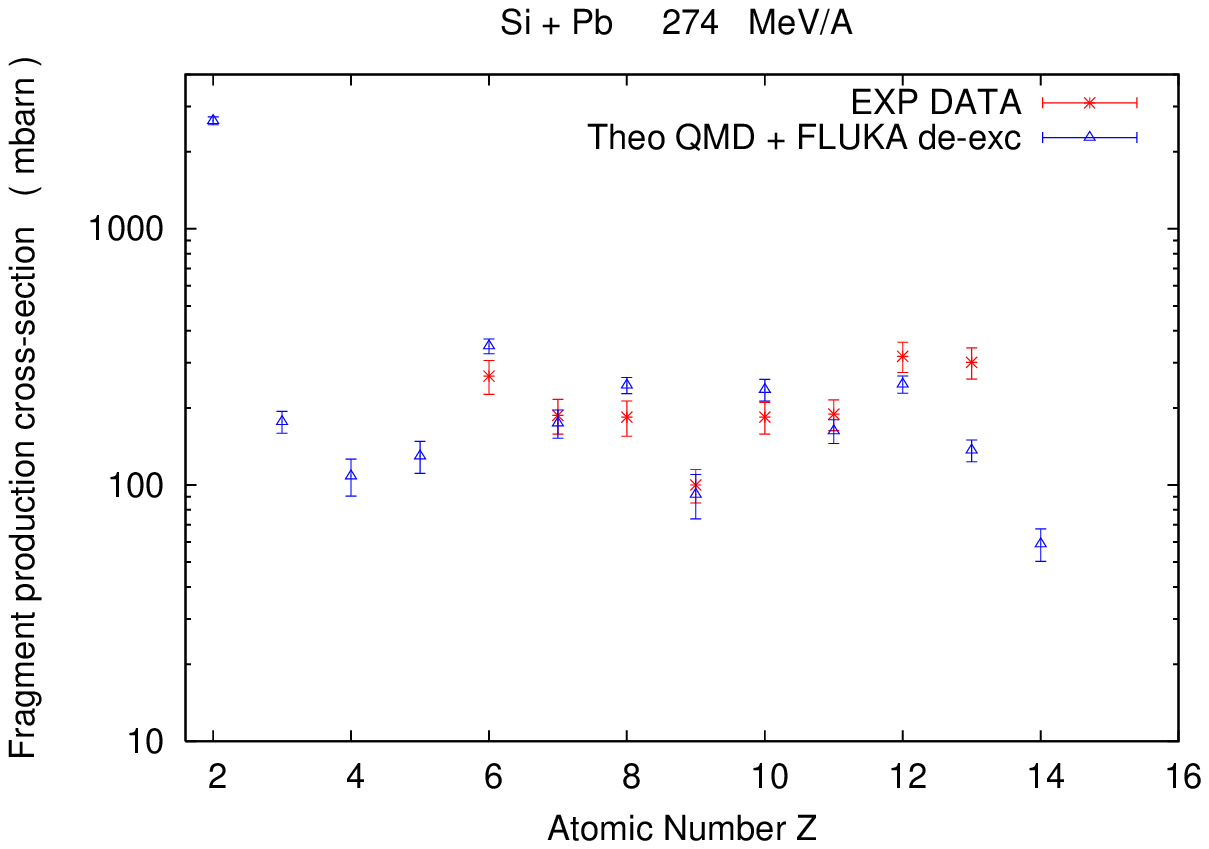}
}
\resizebox{0.815\columnwidth}{!}{%
   \includegraphics[bb=50 50 400 280, clip]{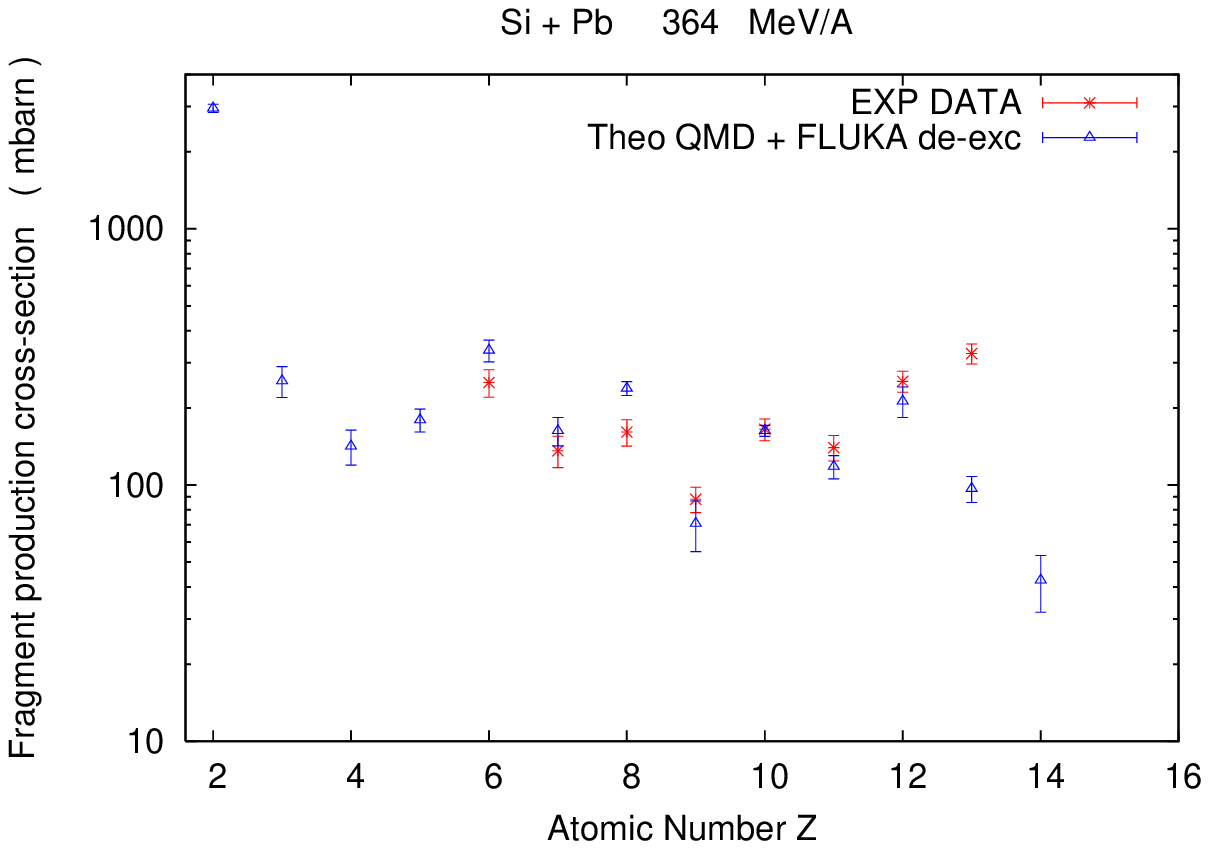}  
}
\resizebox{0.815\columnwidth}{!}{%
   \includegraphics[bb=50 50 400 280, clip]{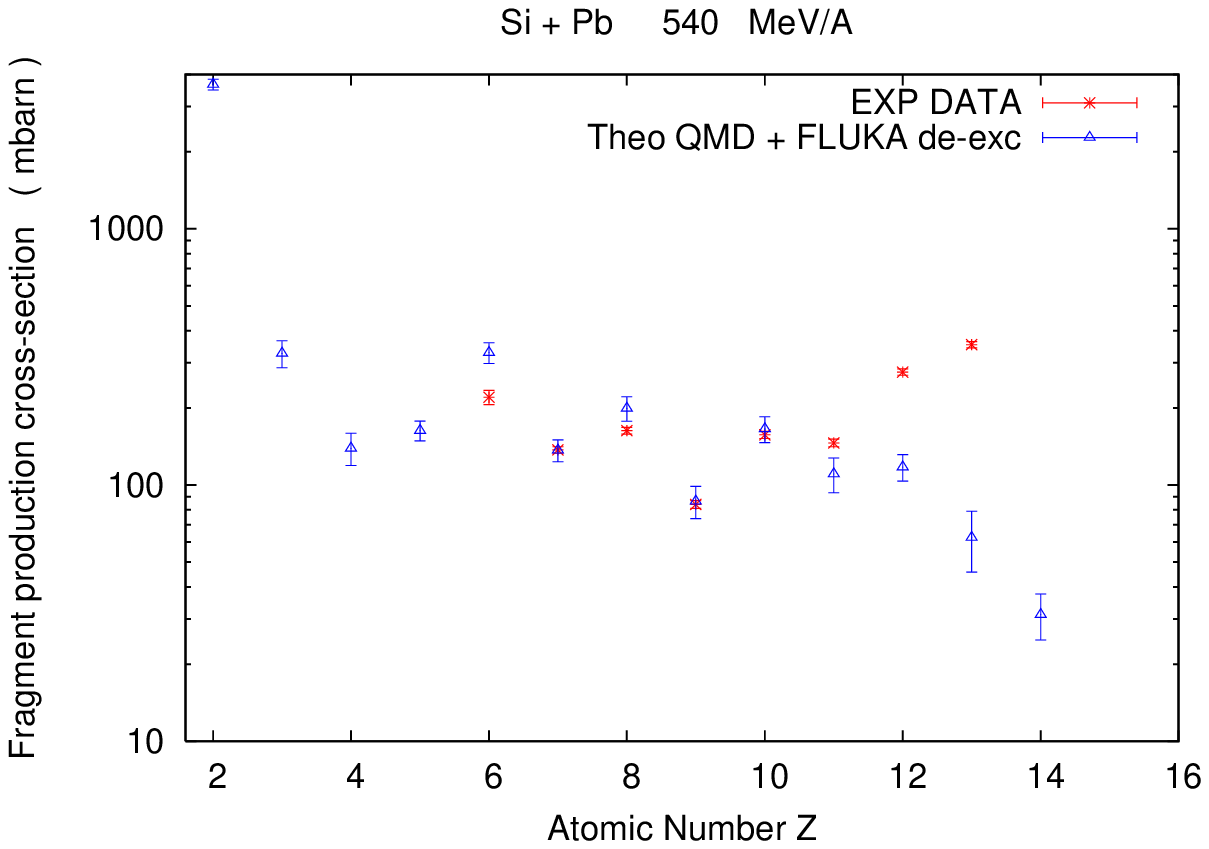}  
}
\caption{The same as figs~\ref{fig:4} and~\ref{fig:5}
for Si ions impinging on Pb at 274, 364 and 540 MeV/A 
bombarding energies, respectively in the upper, 
intermediate and lower panel.
\label{fig:6}}
\end{figure}
\begin{figure}[t]
\centering
\resizebox{0.77\columnwidth}{!}{%
   \includegraphics[bb=52 50 405 285, clip]{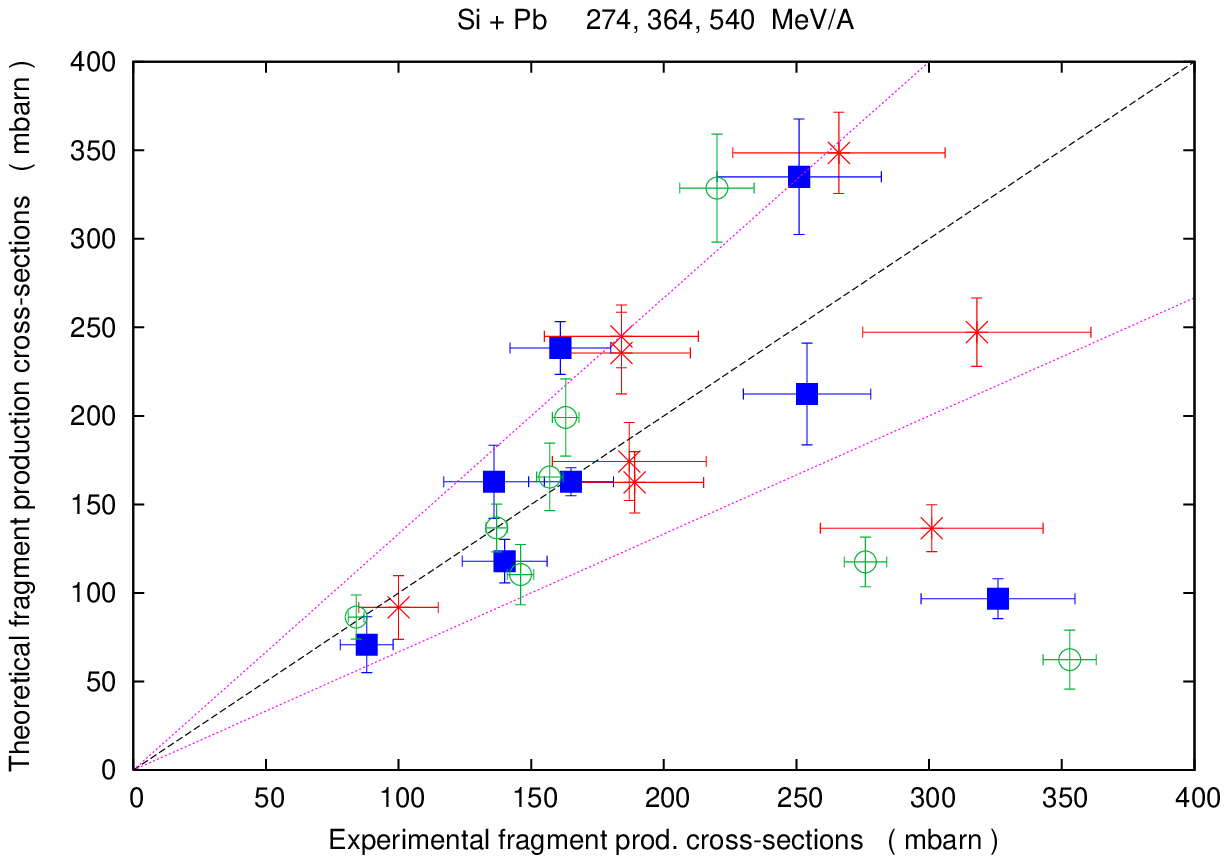}  
}
\caption{Scatterplot of fragment production cross-sections predicted by
QMD + FLUKA de-excitation (vertical axis) vs. measured 
cross-sections (horizontal axis) 
for Si ions impinging on Pb at 274 (crosses), 364 (squares) 
and 540 (circles) MeV/A bombarding enegies. The errorbars on the
theoretical results are statistical only.
The two lines respectively above and below the diagonal 
correspond to deviations of the
modelled results from the experimental ones by $\pm$ 33 \%. 
The largest discrepancies are seen for the Z = 13 fragment production
cross-sections.}
\label{fig:7}
\end{figure}
\subsection{Neutron emission}
As far as neutron emission is concerned, 
performances of the model in predicting double-differential neutron production
cross-sections in the projectile energy range from several tens MeV/A
up to $\sim$ 400 MeV/A have already been shown 
in refs.~\cite{garzvar06,garzrila06,cosparmv}.
The QMD we have developed so far includes non-relativistic potentials,
thus it is expected to work at energies up to a few hundreds MeV/A,
limited also by the inability to deal with pion production and reinteraction. 
In the effort of better understanding up to which energies it can be
reasonably applied, and which modifications have to be included to improve
it, a few simulations at energies above $\sim$ 500 MeV/A have
also been performed. As an example, the neutron double-differential
production cross section for Ar + C at 560 MeV/A bombarding energy
is shown in fig.~\ref{fig:1}. 
The experimental data taken from ref.~\cite{iwata} are plotted
together with the theoretical curves. 
It is apparent that, as far as forward emission angles are
concerned (5 -- 10$^{\mathrm{o}}$), the model underestimates neutron emission
tails from $\sim$ 800 MeV/A up to the highest energies.
On the other hand, at $> 20^{\mathrm{o}}$ emission angles,  the agreement 
of the theoretical results with the experimental data is quite
encouraging, and at angles $\ge 40^{\mathrm{o}}$
the neutron tails are nicely reproduced. 
\subsection{Proton emission}
While systematic data concerning neutron emission in heavy-ion collisions
exist, as far as proton emission in ion-ion collisions 
is concerned, less data are available in literature. 
We thus start considering nucleon induced reactions, for which
many more data concerning proton emission exist.
Indication of the performances of our code in reproducing
the experimental data presented by \cite{proto} is shown 
in fig.~\ref{fig:2} and~\ref{fig:3} 
for p + C at 300 and 392 MeV bombarding energies, 
respectively. 
The linear scale on the energy axis allows to better appreciate
the position of the broad emission peaks.
With the exception of the quasi-elastic peak,
the general agreement is quite good as far as forward emission
angles are concerned, while the neutron 
tails at larger angles are underestimated,
especially at the highest energies.
The slight departure from smoothness of some of 
the lines obtained by simulation
is not due to statistics. It is instead an artifact introduced
by the fact that quite a few QMD ion initial states
have been used to perform these simulations, where projectiles are
single protons. In fact, only seven $^\mathrm{12}$C 
initial states have been used 
to obtain these figures. As a general result, 
when nucleon - ion or light ion - light ion simulations
are performed, better results are obtained considering a larger
number of QMD initial states.     
\subsection{Projectile-like fragment emission}
As far as fragment emission is concerned, a few results obtained
by our QMD coupled to FLUKA de-excitation have already been presented 
in refs.~\cite{garzrila06,cosparmv}. 

In the present paper, the focus is
given to the case of projectile-like fragments produced by Si ions
impinging on targets made of C, Cu and Pb.
Systematic studies of Si fragmentation on targets of
light, intermediate and heavy composition
have been carried out in the last few years  
at the HIMAC in Chiba, and at the BNL AGS. The results of these experiments
were recently published~\cite{silicio}. Comparisons with the predictions
of our model, as far as projectile-like 
fragment production cross-sections are concerned,
are under way.
A few results, for the lowest energies cases, are plotted in 
figs.~\ref{fig:4},~\ref{fig:5},~\ref{fig:6}.
Detector acceptance was taken into account in performing our simulations
by means of cuts in energy and angle, 
which allow to select only projectile-like fragments escaping the target 
within a few degrees around forward direction, according to the 
angular acceptances given in~\cite{silicio}.  
Due to the features of the detectors used in
these experiments, data are available 
for fragments with charge Z = 6 - 13 only. For lower charges
(Z~$\le$~5),
it was very difficult to disentangle the peaks corresponding to the separate
contribution of each fragment species to the energy released in the
Si detectors used. 
Sometime, also the broad peaks corresponding to the Z = 6, 7
charges were determined with some uncertainties, especially at the lowest
energies.  
Anyway, the authors of ref.~\cite{silicio} 
observe that their largest uncertainties
concern the determination of the Z~=~13 fragment production
cross-sections, due to  
superposition effects
with the high-energy tails of primary  ions that cross 
the targets without interacting, which have to be cut. These uncertainties
are more important at lower energies, due to the use of 
thin targets to avoid significant energy losses of projectiles, which
would lead to misleading reaction cross-sections.    

In all cases discussed in the present paper, 
the theo\-re\-ti\-cal model systematically
underestimates the Z~=~13 cross-sections, while for 
Cu and Pb targets it overestimates the Z~=~6 production cross-sections.
The best agreement with the experimental
data is instead observed for fragments with Z~=~10,~11,
and for fragments with Z~=~12 at energies below 400 MeV/A. 
At higher energies, an underestimation of the Z~=~12 fragment
abundances occurs as well. 
This could be an indication that
the QMD developed so far 
gives more reliable results when used in the study of central
collisions, while it is more difficult to apply it to the study of
stripping reactions, where only one or two nucleons of the projectile ion
interact with the target.
Furthermore, the authors of~ref.~\cite{silicio}
suggest the hyphotesis that electromagnetic dissociation can contribute
to the Z = 12,~13 fragment production cross-sections at high energy
for the heaviest targets. At present, we have not tested this hypothesis
yet. Finally,
as far as Fluorine fragments (Z~=~9) are concerned, in all cases
the theoretical model
confirms their suppression, at least from a qualitative point of view, 
with respect to the abundance
of the other species close in charge, 
as observed in the experiment. 

To obtain a global overview of 
the deviations of the predictions of our theoretical model from
the experiments, the results  
can be summarized in scatterplots, such as the one shown
in fig~\ref{fig:7} for the Si + Pb case, including data at 
different energies for fragments with Z = 6 - 13.
With the exception of the cases of the heaviest (and lightest) 
charged fragments already discussed, 
the figure does not show particular systematics.
For the heavy target considered, it is apparent that 
the discrepancies between the model
predictions and the experimental data do not exceed 
$\sim$ 30 - 35\% in most cases.
Theoretical fragmentation models based on QMD codes
provide, in general, more reliable predictions in case of intermediate 
and heavy mass targets than in case of~lighter~ones.  

\vspace{-1mm}
\begin{acknowledgement}
Collaboration with F. Ballarini, G. Battistoni, F. Cerutti, A. Fass\`o,
A.~Ferrari, E. Gadioli, A. Ottolenghi, L.S. Pinsky, J. Ranft and P.R.~Sala
is acknowledeged. This work was supported by the University of Milano.
\end{acknowledgement}

\end{document}